# Analysis of Explainable Artificial Intelligence Methods on Medical Image Classification

Joy Purohit
Department Of Computer Engineering
and Information Technology
Veermata Jijabai Technological Institute
Mumbai, India
purohitjoy21@gmail.com

Ishaan Shivhare
Department Of Computer Engineering
and Information Technology
Veermata Jijabai Technological Institute
Mumbai, India
ishaanshivhare2001@gmail.com

Vinay Jogani
Department Of Computer Engineering
and Information Technology
Veermata Jijabai Technological Institute
Mumbai, India
joganivinay@gmail.com

Seema C Shrawne
Department Of Computer Engineering
and Information Technology
Veermata Jijabai Technological Institute
Mumbai, India
scshrawne@ce.vjti.ac.in

*Abstract*— The use of deep learning in computer vision tasks such as image classification has led to a rapid increase in the performance of such systems. Due to this substantial increment in the utility of these systems, the use of artificial intelligence in many critical tasks has exploded. In the medical domain, medical image classification systems are being adopted due to their high accuracy and near parity with human physicians in many tasks. However, these artificial intelligence systems are extremely complex and are considered 'black boxes' by scientists, due to the difficulty in interpreting what exactly led to the predictions made by these models. When these systems are being used to assist high-stakes decision-making, it is extremely important to be able to understand, verify and justify the conclusions reached by the model. The research techniques being used to gain insight into the black-box models are in the field of explainable artificial intelligence (XAI). In this paper, we evaluated three different XAI methods across two convolutional neural network models trained to classify lung cancer from histopathological images. We visualized the outputs and analyzed the performance of these methods, in order to better understand how to apply explainable artificial intelligence in the medical domain.

*Keywords—Explainable Artificial Intelligence, Gradient-weighted Class Activation Mapping, Integrated Gradient, Local interpretable model-agnostic explanations, interpretable deep learning, model transparency.*

I. INTRODUCTION

The advent of deep learning techniques has majorly contributed to the rapid progress in the field of computer vision. Previously, computer vision tasks were solely conducted using human-designed systems. For instance, a statistical classifier that used manually selected features of an image could be an essential component of the system. These features included low-level image properties such as corners or edges as well as features that were high-level like the speculated border of cancer. In contrast, deep learning systems learn these features automatically with the help of deep neural networks. The substantial increase in the usefulness and accuracy of medical image analysis systems has led to a greater level of integration in the healthcare industry [15].

Machine learning systems have achieved diagnostic parity with physicians on tasks such as radiology, ophthalmology and dermatology [14]. When applied to such critical tasks, it is extremely important for the artificial intelligence model to not only provide a decision or a diagnosis but also to explain the process behind such decisions. In contrast to simpler and self-explaining models such as linear regression, deep neural networks lack interpretability due to their complex design. Neural networks typically consist of multiple layers connected via many nonlinear interlaced relations. Due to the nature of this design, even on the examination of these layers and all their links, it would be very difficult to fully comprehend what led to the decision made by the neural network. From a scientific perspective, these systems are black-box artificial intelligence solutions that have very little transparency and interpretability.

Recently, substantial research has been conducted on techniques to better understand these black-box models. This research falls under the domain of explainable artificial intelligence (XAI) [1]. These methods help evaluate the models beyond standard performance metrics and help understand them better by examining their explanations [2]. Beyond helping to justify the decisions made by these models, XAI may also prove to be a helpful tool to learn new facts, and possibly assist in discovering new things pertaining to the task that had not been observed before. In this study, we apply three different XAI methods to two different medical image classification models, both of which have been trained to identify lung cancer from histopathological images. We then visualize the output of the XAI methods and compare their performance across the two models. Thus we are able to evaluate the utility of these methods in the context of medical image classification systems.

II. RELATED WORK

Samek et.al [3] explains the need for Explainable Artificial Intelligence and also highlights the necessity of the same. Machine learning and deep learning models have become so advanced and complicated that the decisions made by them using black box techniques provide no information on why the decision was made. Therefore to find out the reasons why the model came to that particular result XAI was introduced. Prior to XAI, these models lacked transparency. The paper talks about methods that can help determine why the decisions were taken by the models. Jin et.al [4] and Antoniadi et.al [12] discuss the need for proper evaluation criteria for XAI methods in the field of medicine. The paper emphasizes that not only does the explanation technique need to be technical, but also easy to understand for someone working in the medical field. To counter the gap in explanation the paper conducts an evaluation on a novel problem of multimodal medical image explanation with 2 tasks and proposes new evaluation metrics to explain the results of the classification which can be technically sound and also make sense to a person from the medical field.

## III. EXPLAINABLE ARTIFICIAL INTELLIGENCE

Explainable artificial intelligence methods that are used to explain predictions made by models can be classified into two classes [11]:

- Model-Agnostic: mostly suited to post-hoc analysis and not constrained to particular model architecture. The internal model weights and structural parameters are not available to these approaches.. LIME is a model-agnostic method.

- Model-Specific: dependance on the specific model's parameters. Integrated Gradients and Grad-Cam are model specific methods.

### A. Grad-Cam

Grad-CAM [5] works on the principle of gradients, where these gradients are computed for each channel of an output feature map with respect to the predicted class of an input image. These values obtained for each channel are assigned as weights to the channels respectively, that is, the higher the gradient obtained for a channel- the more the importance of the channel in the prediction of the class. This allows highlighting pixels in an image which are contributing significantly to the model predictions.

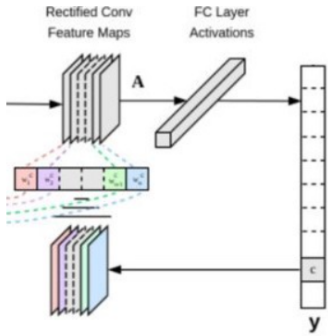

Figure 1: Explanation of Grad-CAM

As illustrated in figure 1, the feature maps obtained from the last convolution layer are utilized to compute the Grad-CAM heatmaps to preserve spatial local information of the object in an input image.

$$a^k_c = 1/Z \sum_i \sum_j (\delta Y^c / \delta A^k_{ij}) \quad (1)$$

To obtain the importance for each feature map 'k', the gradient is computed of the top predicted class 'c' with respect to the feature maps of the last convolution layer 'Ak' i.e. $\partial y^c / \partial A^k$. These computed gradients then go under global average pooling along the width and height dimensions (indexed by i & j in equation 1) which outputs the importance of feature map k with respect to target class c.

### B. Integrated Gradient

Integrated Gradient [6] is an XAI model that works toward computing the gradient of the model's predictions resulting from its input features. Integrated Gradient can be applied to all kinds of data. The method helps understand feature weightage by extracting the rules of the network. Integrated Gradient is built upon two axioms, one being sensitivity and the other being implementation invariance. We first take an image as a baseline image, which is an input image utilised as a starting point for determining feature importance, to calculate the sensitivity. Then after that we create a series of images that we interpolate from the reference picture to the real picture to determine the integrated gradients (IG). When the attributions on the actual image and the baseline image are the same, implementation invariance is satisfied.

The integrated gradient along the $i^{th}$ dimension for an input x and baseline x′ is defined in equation (2). Here, $\partial F(x)$ is $\partial F(x)/\partial x_i$, which is the gradient of F(x) along the $i^{th}$ dimension.

$$IG(x) ::= (x_i - x')=\int_0^1 [\partial F(x'+\alpha \times (x-x'))/\partial x]d \quad (2)$$

Marco et.al [13] highlights how Integrated Gradient is a better attribution method in DNN, demonstrating that it is an extremely useful XAI method that could be used to help explain predictions made by the models we further experiment upon.

### C. Lime

Local Interpretable Model-agnostic Explanations [2] (LIME) is a technique used to explain a machine learning model's individual predictions because of its local character. It is a method of visualization that aids in the explanation of certain predictions. It can be used with any supervised regression or classification model because it is model-independent. Marco Tulio Ribeiro, Sameer Singh, and Carlos Guestrin presented LIME in 2016 [2]. LIME operates under the presumption that all complicated models are linear on a local scale. It attempts to fit a straightforward model to a single observation that will replicate the local behavior of the global model. The predictions of the more sophisticated model can then be locally explained using the simple model.

The core notion of Lime is that the surrogate model is trained on various interpretable features zz from the original model inputs xx. As seen in equation 3, comprehensible illustration h(Ž) = X where h is a mapping between the binary vectors and the related altered picture. Ž is a binary vector denoting the presence or absence of a continuous region of the image. Each produced sample is weighted by the corresponding similarity of X̌ to the instance of interest X, which is an intriguing feature of the Lime technique. In our instance, we employed an exponential kernel input $\pi_x(2) = \exp(-||x - h(Ž)||^2_F)$ between the original and perturbed inputs as a similarity function. The surrogate model is created by the reduction of

$$g = \text{argmin}_{g'} \sum_{\check{z}} \pi_x(\check{Z}) \cdot ((f_c \circ h)(\check{Z}) - g'(\check{Z})) \quad (3)$$

## IV. MEDICAL IMAGE CLASSIFICATION

Deep Neural Networks (DNN) has become one of the most effective machine learning techniques for the task of classifying medical images, which will help medical professionals and enhance diagnosis utilizing these images. In this paper, we will be performing our experiments on two imagenet [9] pre-trained CNN models which are further fine-tuned, training them to classify lung cancer using histopathological images for the same.

### A. Dataset

The models mentioned above are fine-tuned and tested against the 'Lung and Colon Cancer Histopathological Image Dataset (LC25000)' [10] dataset. This dataset comprises 25,000 histopathological images with 5 classes- 3 classes for lung cancer, and 2 classes for colon cancer. All the images in this dataset have dimensions of 768 x 768 pixels and are in the file format 'jpeg'.

These images, which include 750 total images of lung tissue and 500 total images of colon tissue are further augmented using the Python programming language's Augmentor module, which was increased to 25,000. For augmentation purposes, left and right rotations and horizontal and vertical flips were used. In this work, we have used the lung images from this dataset which is a total of 15,000 images in number consisting of 3 classes.

## B. Models

In this paper, two CNN models namely VGG-16 [7] and ResNet-50 [8] having weights pre-trained on Imagenet [9] have been used. Further, these models are fine-tuned on the lung images extracted from the dataset [10] mentioned above. The images from the dataset [10] are resized to 224 x 224 x 3 pixels to train and test the model. The train-test split has been considered as 80-20 in this experiment, and the number of epochs is set to be 5.

Table 1: Accuracy obtained on the test set for the two models.

|  | VGG-16 | Res-Net |
|---|---|---|
| Accuracy | 97.63 % | 98.53 % |

As seen in table 1, these models are effectively able to classify these lung cancer images, hence, achieving good accuracy. This is very important, as the better the model is capable of identifying the correct label, the more effective will be the XAI methods, as these methods leverage the learned representations and features of the model to generate interpretations and explanations of the target images.

## V. METHODOLOGY

In order to understand the usefulness of XAI methods in the application of lung cancer classification, we have performed the implementation of three XAI methods [2,5,6] mentioned above on VGG-16 [7] and ResNet-50 [8] respectively which are trained to classify lung cancer images.

First of all, we train the two CNN models enabling them to effectively classify lung cancer images. Following that, we apply Grad-cam [5], Integrated gradient [6], and LIME [2] respectively on both models to visualize the feature interpretability of these methods. In addition, an average computation time per image is computed for each method. The process is described in greater detail below.

### A. Grad-Cam

Grad-CAM is implemented in accordance with equation 1. The steps of implementation are stated as follows:

1. First, we build a model that associates the input image with both the activations and predictions of the final convolution layer.

2. Second, we calculate the gradient between the top predicted class and the activations of the final convolution layer for the input image.

3. Thirdly, global average pooling is conducted on the gradient of the output neuron with respect to the output feature maps of the last convolution layer

4. Then every channel of the output feature map of the last convolution layer is multiplied by its importance (calculated in the third step). In order to generate a heatmap class activation for visualization, all the channels are added together.

5. The heatmap generated in the step above is then superimposed on the original image to identify the critical regions contributing to the classification of the image.

### B. Integrated Gradient

Integrated gradient is implemented in accordance with equation 2. We start with a baseline image, which represents the impact of the absence of each pixel on the "Lung cancer image" prediction to contrast with the impact of each pixel on the "Lung cancer image" prediction when present in the input image. In this experiment, the baseline is taken as a tensor of size 224 x 224 x 3 containing zeros ( black image). The steps of implementation are stated as follows:

1. First of all, linear interpolations are generated starting from the baseline till the input image. This is done using '50' intervals where each interval is a step in the feature space towards the input image and away from the baseline.

2. Compute gradients between model output predictions with respect to input features. This provides information about the pixels which have the strongest effect on the class probabilities predicted by the model. This gradient is calculated for each interpolation and further all gradients are concatenated into one tensor.

3. Integral approximation is performed which is averaging the gradients of 50 gradients obtained at 50 steps while generating linear interpolations as described in step 1. Basically, the gradients in the concatenated tensor obtained in step 3 are averaged.

4. Now these integrated gradients are scaled with respect to the original image. This ensures that the values obtained from the multiple interpolated images are in the same scale (avoiding unfair advantage) and is an accurate representation of the pixel importance.

5. Now these integrated gradients are visualized using an attribution mask and an overlay of this attribution mask (overlay strength=0.4) on the original input image.

### C. Lime

The major parameters used for implementing LIME are as follows:

1. Input image - image for which LIME explanations are wanted.

2. Model predictions for the input image.

3. top_labels= the number of labels that is expected of LIME to show. This experiment has tob_labels=3.

4. num_samples=to calculate the number of synthetic data points that LIME will produce that are identical to our input. In this experiment, num_samples=1000.

## VI. RESULTS AND ANALYSIS

### A. Interpretability of Grad-Cam

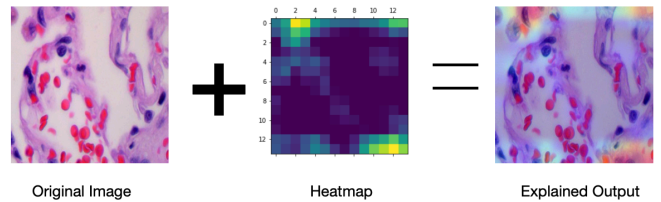

Figure 2: Explanations generated on VGG-16

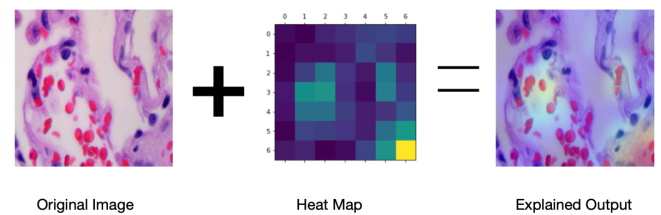

Figure 3: Explanations generated on Resnet-50

Figures 2 and 3 illustrate the process of generating explanations from an input image where a heatmap is generated by the GradCam method. This heatmap is indicative of the relationship between the features of the image and the prediction made by the model. The heatmap highlights the important regions that contributed to the final prediction. The higher the saturation, the higher the weightage placed on those pixels. Further, the heatmap is then scaled to and superimposed upon the original image to obtain explainable images shown on the right for figures 2 and 3. This provides a concrete visualization of the important features (pixels) of the image that the model considered most critical to its prediction.

### B. Interpretability of Integrated Gradient

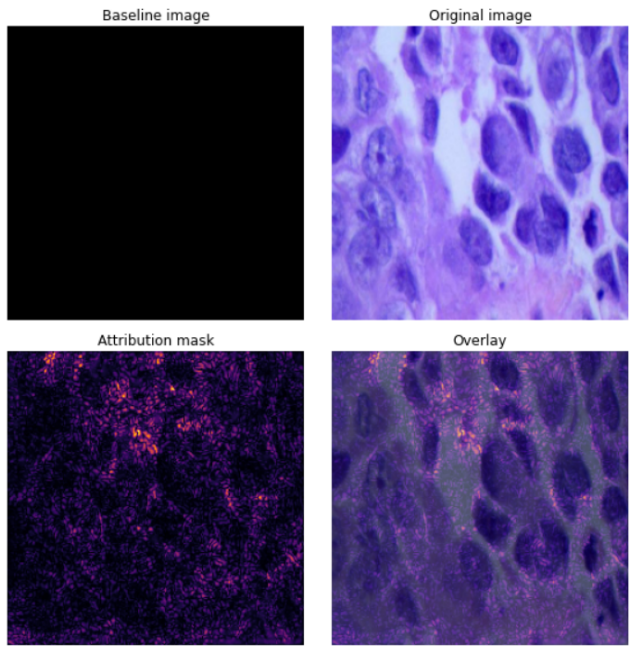

Figure 4: Representations obtained on VGG-16

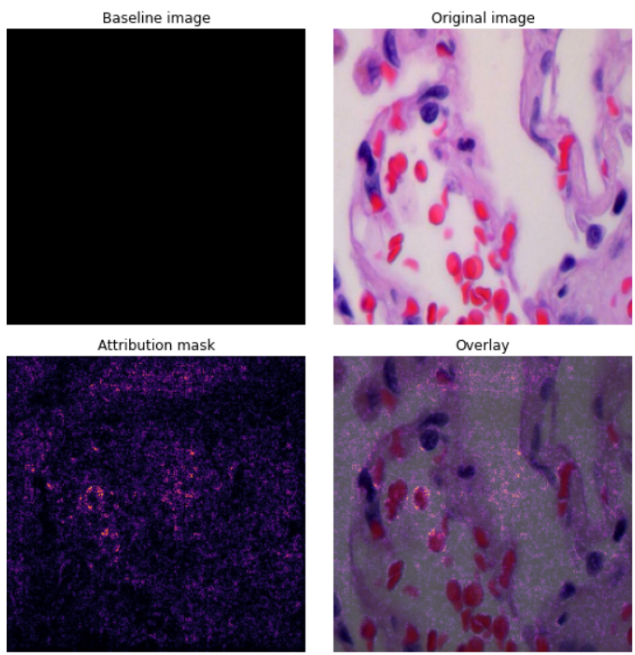

Figure 5: Representations obtained on ResNet-50

Figures 4 and 5 illustrate the visualizations of the interpretability provided by the integrated gradient method for a test lung cancer image. The attribution mask highlights all of the important pixels on the baseline image, with the intensity of the highlighting increasing as the pixel's contribution to model prediction increases. This provides readers with information about all the critical regions of the histopathological image that are cancerous. "Overlay" is a superimposition of the attribution mask on the original image to visualize the critical pixels on the original image, which greatly contributes to the model predictions.

### C. Interpretability of Lime

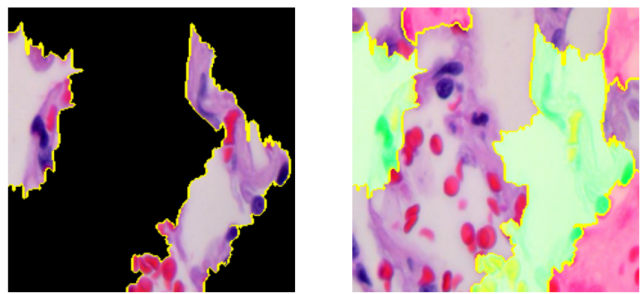

6 (a)     6 (b)

Figure 6: Output explanations on VGG-16

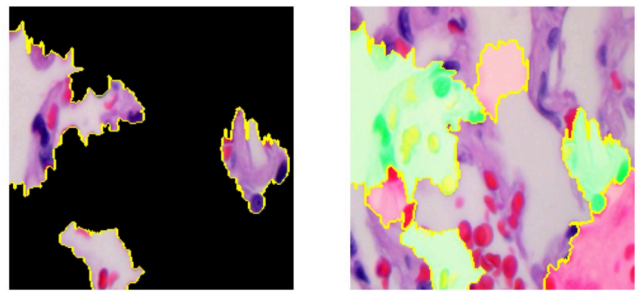

7 (a)     7 (b)

Figure 7: Output explanations on ResNet-50

Figures 6 and 7 describe the interpretations drawn of the input image from the LIME method. (6a,7a) takes all the important regions from the original image contributing to the model predictions and visualizes on a baseline image isolating the regions of importance drawn from the original image. Further, in (6b,7b) the green regions determine the pixels which increase the probability of the class predicted by the model. However, the pixels marked in red suppress the probability of the image being classified according to the model predictions.

By highlighting the regions of importance, these methods are thus able to aid human physicians in verifying a predicted diagnosis by considering the features that the model places the most weight on to reach its decision. Further, it helps explain the reasoning of the prediction made by the medical image classification system, which is important in such cases where high-stakes decisions are made.

### D. Comparative Anaylsis

We further conducted a comparative analysis of the three XAI methods across the two models on the basis of their computational time requirements. The results obtained for the same can be seen in table 1 below.

Table 1: Average Computation time (in seconds) for each model calculated for the testing set compared across models

|  | Grad-Cam | Integrated Gradients | LIME |
|---|---|---|---|
| VGG-16 | 0.037 | 0.071 | 15.47 |
| ResNet-50 | 1.54 | 6.46 | 15.72 |

Observations recorded in Table 1 inform that the fastest explainable method is Grad-cam whereas the slowest one is LIME. Further, it can be seen in table 1 that these 3 methods perform faster on the VGG-16 model as compared to the ResNet-50 model as the number of parameters in the ResNet-50 architecture is greater, thus making it a denser model. Due to this difference in the complexity of the model, it takes more time for the XAI methods to generate explanations for predictions made by ResNet-50.

This huge difference in computation time between LIME and the other two methods is due to the fact that LIME is a model agnostic method. Hence, the method does not have access to the gradients of the trained model. The process of generating these explanations entails sampling and obtaining a surrogate dataset. In order to do so, it produces 5000 samples of the feature vector that follow the normal distribution and obtains target variables for each of these samples using the model to predict it. This, along with the employment of a Ridge Regression model in its procedure contributes to the length of time it takes to generate an example. Hence when a large number of explanations are to be generated, the other XAI methods Integrated Gradients and GradCam may prove to be more useful. However, in cases where a model specific method may not be permissible due to a lack of access to the gradients of the model whose predictions are to be explained, this trade-off for increased computation time may have to be made.

## VII. CONCLUSION

In this study, we compared three XAI approaches for deciphering black-box predictions produced from cutting-edge medical image classification algorithms trained to classify lung cancer on histopathological images. In order to quantify the performance of the XAI methods, we compared the average computation time taken to generate the explanations required to visualize the results across the different methods. We also compared the explanations of each model by each method.

## VIII. FUTURE SCOPE

The objective of the research conducted in this paper was to analyze methods to introduce explainability in the medical domain, specifically for highly accurate medical image classification systems. The field of XAI is extremely useful in order to build trustworthiness in these systems. In order to better understand the most appropriate XAI approach that is to be taken in different use-cases, we can extend the scope of the paper in the future by exploring the effectiveness of these methods on different datasets, and use better quantitative metrics to compare these methods with each other [17]. Further, more techniques to explain model predictions can also be explored, along with the discussion of techniques to introduce explainability in the model itself [16].